# Implementation of Technology Acceptance Model (TAM) and Importance Performance Analysis (IPA) in Testing the Ease and Usability of E-wallet Applications

Dedi Saputra[1], Burcu Gürbüz[2*]

[1] Department of Informatics and Technology, Bina Sarana Informatika University, Jakarta, Indonesia,
[2] Institute of Mathematics, Johannes Gutenberg-University Mainz, Germany,
[*] Corresponding author: burcu.gurbuz@uni-mainz.de

**Abstract**

Digital payment innovation is currently increasingly needed by the community, especially in making non-cash payment transactions. The purpose of this research is to know and measure the ease and usefulness of e-wallet digital wallet services, especially in the GoPay application. The population in this study are users of the Go-Pay service on the GO-JEK platform. The sample of this study consisted of 124 respondents from distributing questionnaires in Depok, West Java using a modified Technology Acceptance Model (TAM) based on existing references. The data processing in this research uses Importance Performance Analysis (IPA) analysis. The results show that based on the gap analysis, it is found that in general Go-Pay users are not satisfied with the current service quality. Based on the IPA analysis, the priority scale of E-Wallet Go-Pay quality improvement can be mapped, where quadrant I is the highest priority scale according to the user's perspective: [1], [4], [5], and [6]. These three items must be upgraded immediately by the manager to meet user expectations. Areas that become the achievements or advantages of the GoPay E-Wallet that must be maintained are in quadrant II, namely: [2] and [3]. From this explanation, it can be concluded that in general the E-Wallet GoPay Service must be improved to improve its service performance.

**Keywords** : E-Wallet Service, TAM Model, IPA Analysis

## 1. Introduction

The increase in digital technology was also followed by the increase in digital payment technology. The use of Digital Technology in making payment transactions in the form of electronic money (E-Wallet) or digital wallets (e-wallets) is a necessity in the midst of the current Covid-19 epidemic, especially in terms of avoiding physical contact (physical distancing) and transactions. in cash (cash). E-wallet is an application that allows its users to carry out buying and selling activities online, or other financial service activities. Regarding the Non-Cash payment system, on 14 August 2014 Bank Indonesia launched the National Non-Cash Movement (GNNT) which aims to increase public awareness of the use of non-cash instruments, so that a community or society that uses less cash is gradually formed. Society / LCS) especially in conducting transactions for its economic activities (Tirta Segara, 2014).

LCS transactions in the form of e-wallet application is a transaction application that is needed by the public today, especially in the midst of the Covid-19 pandemic. According to Fatoni at.al, the real impact of the covid-19 outbreak on the use of e-wallets is that the use of online shopping apps is soaring by 300 percent (Nuha, Qomar, & Maulana, 2020). "The electronic wallet (E-wallet) provides all of the functions of today's wallet on one convenient smart card eliminating the need for several cards" (Upadhayaya, 2012). With an e-wallet application, the efficiency of transactions using smart cards can be realized. E-wallet is also a digital payment transaction application that has a high level of security, so many people switch to using the app.

With the increasing use of the internet in Indonesia, there are several business activities to utilize electronic media, one of which is the E-Wallet application that has developed in Indonesia. Based on the data for the second quarter of 2019 obtained from the top 5 Annie Applications, currently the most active e-wallet applications in each month of transactions are still occupied by local applications, namely Go-Pay, OVO, DANA, LinkAja, and Jenius (Liputan6.com, 2019).

A release from CNBC Indonesia states, in the increasingly fierce competition for digital wallets (e-wallets), customers are trying to be acquired by players, but Go-Pay still stands firmly at the top as the most popular e-wallet in Indonesia(cnbcindonesia, 2019). In Indonesia Go-Pay is a fintech found on the Go-Jek platform which is a mobile payment service (Huwaydi & Persada, 2018).

Today, only a smartphone and an internet connection can make consumers get what they want. For example, by using the E-Wallet application or digital wallet as a purchase transaction tool. By using an E-Wallet, consumers can transfer money to the desired bank account or to another E-Wallet as needed. Consumers are also free to make payments for purchases made anytime and anywhere (Bank Indonesia, 2019). From the number of e-wallets that have been circulating and their growing users, it shows that the interest of people who use E-Wallet is increasingly visible. E-Wallet users receive many benefits and conveniences in fulfilling daily needs. In other words, the services provided by E-Wallet products are very helpful for every user. This attitude is what makes all E-Wallet companies compete in providing the best service to attract market share.

What is very important for an application to be accepted by the public on the e-wallet application is seen from the point of view of its usability and ease. This application can be said to be useful and easy to use if it is tested on all available content. In terms of application testing in terms of usability and convenience, it can be done through the Technology Acceptance Model (TAM) and Importance Performance Analysis (IPA) approaches, in addition to using a questionnaire distributed to users as a predetermined sample.

Technology Acceptance Model (TAM) approach as an acceptance modeling application adapted from Theory Reasoned Action (TRA) developed by Fishbein and Ajzen (1975), namely the theory of action reasoned with one premise that a person's reaction and perception of something, will determine the attitude and behavior of the person. The reaction and perception of information technology users will influence their attitude in acceptance of the technology. One of the factors that can influence it is the perception of the usefulness and ease of use of the information system as an act of reason in the context of technology users, so that the reason a person sees the benefits and ease of use of the information system makes the actions / behavior of the person as a benchmark in the acceptance of a system.

The approach of technology acceptance model has combined the attitude of its users to foster a sense of trust in the technology used. The TAM (Technology Acceptance Model) states that behavioral intension to use is determined by two beliefs namely: first, *perceived usefulness* which is defined as the extent to which a person is confident that using the system will improve his performance. Second, *perceived ease of use* is defined as the extent to which one is convinced that using the system is easy (Supriyati & Cholil, 2017). TAM models have been widely used to test the acceptance of technology by system users in a variety of contexts(Kurniawati, Arif, & Winarno, 2017). TAM's main objective is to provide a basic step from the impact of an external factor on internal beliefs, attitudes and intentions (Kurniawan, Saputra, & Prasetyawan, 2018).

This Importance Performance Analysis (IPA) method was first introduced by Martilla and james (1977) aimed to measure the relationship between consumer perception and priority of improving product/service quality (Fajri, Sugiarto, & Anggraini, 2019). IPA has been generally accepted and used in various fields of study because of the ease to be applied and the display of analysis results that facilitate the proposed performance improvement. Martinez in (Napitupulu, Ariani, & Kadar, 2016). IPA has the main function to display information related to service factors that consumers think greatly influence their satisfaction and loyalty, and service factors that consumers think need to be improved because the current conditions are not satisfactory. IPA combines measurement of importance factor (expectation) and performance level (perception) in a two-dimensional graph that facilitates explanation of data and obtains practical proposals. IPA is a method of measuring customer satisfaction by analyzing the reality felt by customers compared to customer expectations (Lusianti, 2017). There are three analyses used in science, namely depth analysis, gap level analysis, and quadrant analysis (Fatmala & Suprapto, 2018). In this technique, respondents were asked to rate the level of importance and level of performance and then the average value of interest and performance was analyzed in the Importance Performance Matrix, where the x-axis represents performance (perception) while the y-axis represents the interest (expectation). IPA graph is divided into four quadrants based on the results of importance performance measurement that provides interpretation as seen below:

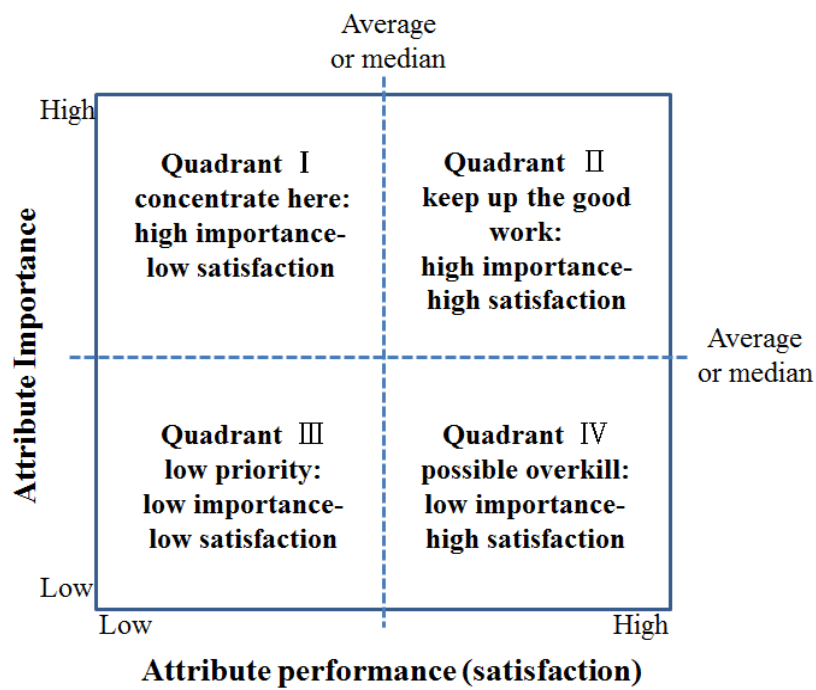

Figure 1. Quadrant Division Importance Performance Analysis

The *Importance-Performance Analysis* (IPA) diagram consists of four quadrants:

1. Quadrant I, a region that contains items with a relatively high level of importance but the reality is not yet in line with user expectations. Items that enter this quadrant must be performance enhanced immediately.
2. Quadrant II, an area containing items that have a relatively high level of importance with a relatively high level of satisfaction. Items that enter this quadrant is considered as a supporting factor for user satisfaction so it must be maintained because all these items make the product or service superior in the eyes of the user.
3. Quadrant III, a region containing items with a relatively low level of importance and the fact that its performance is not very special with a relatively low level of satisfaction. Items that enter this quadrant have very little effect on the benefits felt by the user.
4. Quadrant IV, the territory containing items with a relatively low level of importance and perceived by the user is excessive with a relatively high level of satisfaction. The cost used to support items entering this quadrant can be reduced in order to save expenses.

The issues raised and the focus of this research include: (1) What are the results of the analysis of the usability and ease of the GoPay E-Wallet service based on the TAM and IPA models ?; (2) What are the recommendations that will be submitted to developers based on the results of testing conducted by the author to improve the quality of GoPay E-Wallet services?

Referring to what has been done by previous researchers who have been trusted and published, several studies that the authors can refer to include:
The research entitled Intentions to Use Millennial Generation Electronic Wallets in Three Indonesian "Unicorn" Startups Based on Modification Tam explained that the factors of perceived usefulness, perceived ease of use and subjective norms proved to have a positive effect on the intention to use e-wallets. Thus it can be concluded that: 1) the more a person feels the benefits of e-wallets, the more motivated he is to use e-wallets, 2) the more someone feels the ease of using e-wallets, the more motivated he is to use e-wallets and 3) the more someone feels that the people around him support the use of e-wallets, so he is increasingly using e-wallets(Hutami & Septyarini, 2019).

The research entitled The Effect of Perceptions of Usefulness and Perceptions of Ease of Use on Interest in Using Go-Pay Digital Payment Services concluded that to test the factors that influence the interest in using electronic money using the testing model used in this study is the Technology Acceptance Model (TAM). This research was conducted on users of the GO-PAY digital payment service who are domiciled in Jakarta. The sampling technique used was the convenience sampling technique. Meanwhile, the data obtained is in the form of primary data because data collection is done by distributing online questionnaires. A total of 125 questionnaires were received and all of them can be processed by the researcher. Data analysis was performed using variant-based structural equation analysis (SEM) techniques or Partial Least Squares (PLS). The results obtained are that perceptions of usefulness and perceived ease of use have a positive influence on interest in use, and perceptions of ease of use have a positive influence on perceptions of usefulness (Joan & Sitinjak, 2019).

Based on the description above, the author tries to make research to test the level of ease and usability of the E-Wallet Go-Pay application service using the TAM model and IPA analysis, so that it is expected to be a positive recommendation for application developers, especially in improving application service performance.

## 2. Reseach Methods

### 2.1. Literature Study
Studying the literature that will be used as a theoretical study in this research which is sourced from reference books, journals and articles taken from the internet.

### 2.2. Problem Identification
Identify what issues will be discussed in relation to the usability and ease of E-Wallet Go-Pay services based on the Technology Acceptance Model (TAM) and Importance Performance Analysis (IPA) models.

### 2.3. Compilation of Research Instruments
At this stage the researcher compiles a questionnaire (questionnaire) which will be used to collect data from respondents based on a modified Technology Acceptance Model (TAM) based on existing references in literature studies.

### 2.4. Data Collection
  a. Population and Sample
     This stage is to search for samples based on a predetermined population.
  b. Instrument Development
     This stage is the determination of the research instrument, namely by using a
      questionnaire. The preparation of this questionnaire is divided into two parts, namely the
      identity of the data source and quantitative. Two types of quantitative data are taken,
      where the data are of importance and performance of the E-Wallet Go-Pay. Then arranged in a
      bundle to be distributed to respondents.
  c. Instrument Testing
     The instrument testing stage is carried out by prerequisite testing where testing is carried
      out to test the level of validity and reliability of the instruments that will be used during the
     data collection process.

### 2.5. Data Analysis
Analyze the results of data processing based on the results of research and existing theories with Importance Performance Analysis.

### 2.6. Conclusions and Suggestions
Conclusions are drawn based on data analysis and checked whether they are in accordance with the aims and objectives of the study and suggestions for further research.

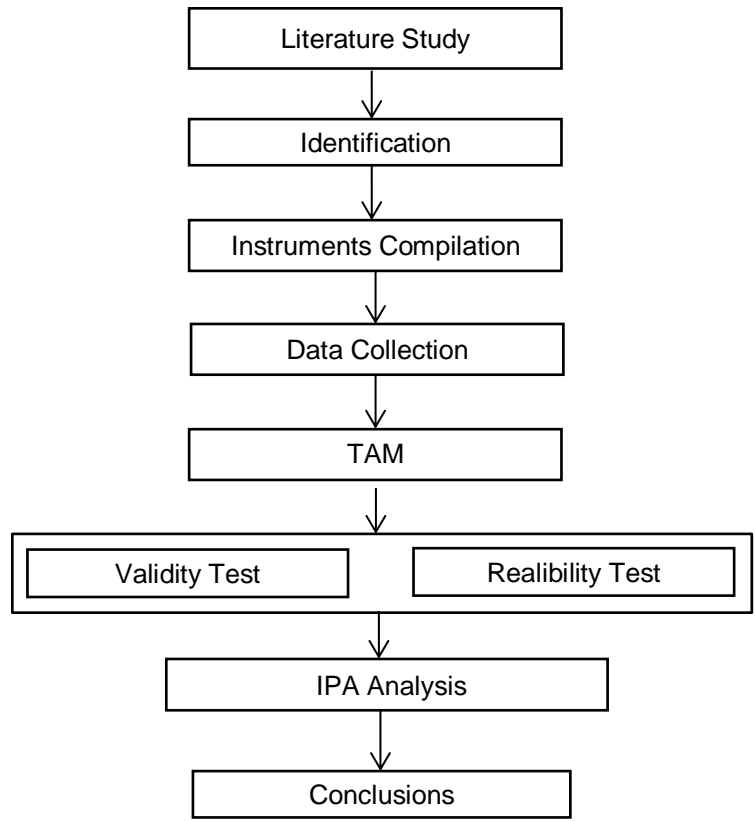

**Figure 2. Research Method**

## 3. Result and Discussion

In this research, the object is the E-Wallet Go-Pay. After the questionnaire data was collected, a demographic analysis of the respondents was carried out, then performed a data quality test (validity and reliability tests) and an analysis of Importance Performance Analysis as the final stage of the study. All tests in this study were carried out with calculations using SPSS 21. To test whether the measuring instrument (instrument) used meets the requirements of a good measuring instrument. So that the resulting data is in accordance with what is being measured.

The purpose of this test is to analyze the usability and convenience of E-Wallet services based on the TAM and IPA model and provide suggestions or recommendations for developers based on the results of these tests in order to improve E-Wallet services. Researchers use a Likert scale to rate:
  a. Perceived Ease of Use / XA1 Importance.
  b. Perceived Usefulness / XA2 Level of Importance (Importance)
  c. Perceived Ease of Use / XB1 Performance Level
  d. Perceived Usefulness / XB2 Performance Level

Here the respondents are asked to provide an assessment of the indicators presented in the questions in the questionnaire. Questions are made in two types, namely the level of importance (importance) and the level of performance (performance). The results of collecting questionnaires that are successfully compiled and suitable for analysis can be shown in the table below:

**Table 1. Questionnaire Collection Results**

| No | Amount | Persentage | Information |
|---|---|---|---|
| **1.** | 93 | 100 % | Complete |
| **2.** | 0 | 0 % | Not Complete |
| **3.** | 0 | 0 % | Not Eligible |

### 3.1. Validity Test

Before testing the validity, the real level is determined (α), namely 5% or 0.05 and the test statistic used is (rho-Spearman), critical value = table value where n = 93. r table = rα; (n-2 ) = r0,05; (91) = 0.206
Validity and reliability tests are carried out only for the Performance Level variable.

#### 3.1.1 Validity Test of Importance Perceived Ease of Use XA1

**Table 2. Results of the Validity Test of the Level of Importance Perceptions of Ease of Use XA1**

| | Correlations | XA11 | XA12 | XA13 | XA14 | XA15 | XA16 | Amount |
|---|---|---|---|---|---|---|---|---|
| XA11 | Pearson Correlation | 1 | ,268** | ,478** | ,467** | ,510** | ,439** | ,738** |
| | Sig. (2-tailed) | | ,009 | ,000 | ,000 | ,000 | ,000 | ,000 |
| | N | 93 | 93 | 93 | 93 | 93 | 93 | 93 |
| XA12 | Pearson Correlation | ,268** | 1 | ,714** | ,463** | ,199 | ,277** | ,690** |
| | Sig. (2-tailed) | ,009 | | ,000 | ,000 | ,056 | ,007 | ,000 |
| | N | 93 | 93 | 93 | 93 | 93 | 93 | 93 |
| XA13 | Pearson Correlation | ,478** | ,714** | 1 | ,521** | ,355** | ,388** | ,811** |
| | Sig. (2-tailed) | ,000 | ,000 | | ,000 | ,000 | ,000 | ,000 |
| | N | 93 | 93 | 93 | 93 | 93 | 93 | 93 |
| XA14 | Pearson Correlation | ,467** | ,463** | ,521** | 1 | ,395** | ,377** | ,751** |
| | Sig. (2-tailed) | ,000 | ,000 | ,000 | | ,000 | ,000 | ,000 |
| | N | 93 | 93 | 93 | 93 | 93 | 93 | 93 |
| XA15 | Pearson Correlation | ,510** | ,199 | ,355** | ,395** | 1 | ,267** | ,636** |
| | Sig. (2-tailed) | ,000 | ,056 | ,000 | ,000 | | ,010 | ,000 |
| | N | 93 | 93 | 93 | 93 | 93 | 93 | 93 |
| XA16 | Pearson Correlation | ,439** | ,277** | ,388** | ,377** | ,267** | 1 | ,644** |
| | Sig. (2-tailed) | ,000 | ,007 | ,000 | ,000 | ,010 | | ,000 |
| | N | 93 | 93 | 93 | 93 | 93 | 93 | 93 |
| Amount | Pearson Correlation | ,738** | ,690** | ,811** | ,751** | ,636** | ,644** | 1 |
| | Sig. (2-tailed) | ,000 | ,000 | ,000 | ,000 | ,000 | ,000 | |
| | N | 93 | 93 | 93 | 93 | 93 | 93 | 93 |

**. Correlation is significant at the 0.01 level (2-tailed).

**Table 3. Comparison of Test Results r Count with r Table of Importance Perceived Ease of Use XA1**

| Variable | r Count | r Table | Decision |
|---|---|---|---|
| **XA11** | 0,738 | 0,206 | *Valid* |
| **XA12** | 0,690 | 0,206 | *Valid* |
| **XA13** | 0,811 | 0,206 | *Valid* |
| **XA14** | 0,751 | 0,206 | *Valid* |
| **XA15** | 0,636 | 0,206 | *Valid* |
| **XA16** | 0,644 | 0,206 | *Valid* |

### 3.1.2. Validity Test of the Level of Importance Perception Ease of Use XA2

**Table 4. Results Validity Test of the Level of Importance Perception Ease of Use XA2**

**Correlations**

| | | XA21 | XA22 | XA23 | XA24 | XA25 | XA26 | Amount |
|---|---|---|---|---|---|---|---|---|
| XA21 | Pearson Correlation | 1 | ,328** | ,442** | ,273** | ,406** | ,009 | ,603** |
| | Sig. (2-tailed) | | ,001 | ,000 | ,008 | ,000 | ,930 | ,000 |
| | N | 93 | 93 | 93 | 93 | 93 | 93 | 93 |
| XA22 | Pearson Correlation | ,328** | 1 | ,333** | ,437** | ,395** | ,227* | ,644** |
| | Sig. (2-tailed) | ,001 | | ,001 | ,000 | ,000 | ,028 | ,000 |
| | N | 93 | 93 | 93 | 93 | 93 | 93 | 93 |
| XA23 | Pearson Correlation | ,442** | ,333** | 1 | ,388** | ,425** | ,171 | ,649** |
| | Sig. (2-tailed) | ,000 | ,001 | | ,000 | ,000 | ,101 | ,000 |
| | N | 93 | 93 | 93 | 93 | 93 | 93 | 93 |
| XA24 | Pearson Correlation | ,273** | ,437** | ,388** | 1 | ,591** | ,818** | ,844** |
| | Sig. (2-tailed) | ,008 | ,000 | ,000 | | ,000 | ,000 | ,000 |
| | N | 93 | 93 | 93 | 93 | 93 | 93 | 93 |
| XA25 | Pearson Correlation | ,406** | ,395** | ,425** | ,591** | 1 | ,428** | ,783** |
| | Sig. (2-tailed) | ,000 | ,000 | ,000 | ,000 | | ,000 | ,000 |
| | N | 93 | 93 | 93 | 93 | 93 | 93 | 93 |
| XA26 | Pearson Correlation | ,009 | ,227* | ,171 | ,818** | ,428** | 1 | ,640** |
| | Sig. (2-tailed) | ,930 | ,028 | ,101 | ,000 | ,000 | | ,000 |
| | N | 93 | 93 | 93 | 93 | 93 | 93 | 93 |
| Jumlah | Pearson Correlation | ,603** | ,644** | ,649** | ,844** | ,783** | ,640** | 1 |
| | Sig. (2-tailed) | ,000 | ,000 | ,000 | ,000 | ,000 | ,000 | |
| | N | 93 | 93 | 93 | 93 | 93 | 93 | 93 |

**. Correlation is significant at the 0.01 level (2-tailed).
*. Correlation is significant at the 0.05 level (2-tailed).

**Table 5. Comparison of Test Results r Count with r Table of Importance Perceived Ease of Use XA2**

| Variable | r Count | r Table | Decision |
|---|---|---|---|
| XA21 | 0,603 | 0,206 | *Valid* |
| XA22 | 0,644 | 0,206 | *Valid* |
| XA23 | 0,649 | 0,206 | *Valid* |
| XA24 | 0,844 | 0,206 | *Valid* |

| | | | | | |
|---|---|---|---|---|---|
| XA25 | 0,783 | 0,206 | Valid |
| XA26 | 0,640 | 0,206 | Valid |

### 3.1.3. Validity Test of the Level of Importance Perception Ease of Use XB1

**Table 6. Results Validity Test of the Level of Importance Perception Ease of Use XB1**

**Correlations**

| | | XB11 | XB12 | XB13 | XB14 | XB15 | XB16 | Amount |
|---|---|---|---|---|---|---|---|---|
| XB11 | Pearson Correlation | 1 | ,292** | ,254* | ,560** | ,543** | ,776** | ,784** |
| | Sig. (2-tailed) | | ,004 | ,014 | ,000 | ,000 | ,000 | ,000 |
| | N | 93 | 93 | 93 | 93 | 93 | 93 | 93 |
| XB12 | Pearson Correlation | ,292** | 1 | ,384** | ,429** | ,194 | ,226* | ,537** |
| | Sig. (2-tailed) | ,004 | | ,000 | ,000 | ,062 | ,029 | ,000 |
| | N | 93 | 93 | 93 | 93 | 93 | 93 | 93 |
| XB13 | Pearson Correlation | ,254* | ,384** | 1 | ,650** | ,251* | ,330** | ,613** |
| | Sig. (2-tailed) | ,014 | ,000 | | ,000 | ,015 | ,001 | ,000 |
| | N | 93 | 93 | 93 | 93 | 93 | 93 | 93 |
| XB14 | Pearson Correlation | ,560** | ,429** | ,650** | 1 | ,666** | ,716** | ,901** |
| | Sig. (2-tailed) | ,000 | ,000 | ,000 | | ,000 | ,000 | ,000 |
| | N | 93 | 93 | 93 | 93 | 93 | 93 | 93 |
| XB15 | Pearson Correlation | ,543** | ,194 | ,251* | ,666** | 1 | ,702** | ,768** |
| | Sig. (2-tailed) | ,000 | ,062 | ,015 | ,000 | | ,000 | ,000 |
| | N | 93 | 93 | 93 | 93 | 93 | 93 | 93 |
| XB16 | Pearson Correlation | ,776** | ,226* | ,330** | ,716** | ,702** | 1 | ,857** |
| | Sig. (2-tailed) | ,000 | ,029 | ,001 | ,000 | ,000 | | ,000 |
| | N | 93 | 93 | 93 | 93 | 93 | 93 | 93 |
| Jumlah | Pearson Correlation | ,784** | ,537** | ,613** | ,901** | ,768** | ,857** | 1 |
| | Sig. (2-tailed) | ,000 | ,000 | ,000 | ,000 | ,000 | ,000 | |
| | N | 93 | 93 | 93 | 93 | 93 | 93 | 93 |

**. Correlation is significant at the 0.01 level (2-tailed).
*. Correlation is significant at the 0.05 level (2-tailed).

**Table 7. Comparison of Test Results r Count with r Table of Importance Perceived Ease of Use XB1**

| Variable | r Count | r Table | Decision |
|---|---|---|---|
| **XB11** | 0,784 | 0,206 | Valid |
| **XB12** | 0,537 | 0,206 | Valid |
| **XB13** | 0,613 | 0,206 | Valid |
| **XB14** | 0,901 | 0,206 | Valid |
| **XB15** | 0,768 | 0,206 | Valid |
| **XB16** | 0,857 | 0,206 | Valid |

### 3.1.4. Validity Test of the Level of Importance Perception Ease of Use XB2

**Table 8. Results Validity Test of the Level of Importance Perception Ease of Use XB2**

| | | XB21 | XB22 | XB23 | XB24 | XB25 | XB26 | Amount |
|---|---|---|---|---|---|---|---|---|
| | | Correlations | | | | | | |
| XB21 | Pearson Correlation | 1 | ,106 | ,082 | -,004 | ,022 | ,028 | ,345** |
| | Sig. (2-tailed) | | ,310 | ,437 | ,973 | ,838 | ,787 | ,001 |
| | N | 93 | 93 | 93 | 93 | 93 | 93 | 93 |
| XB22 | Pearson Correlation | ,106 | 1 | -,033 | ,148 | -,012 | ,081 | ,348** |
| | Sig. (2-tailed) | ,310 | | ,754 | ,157 | ,907 | ,438 | ,001 |
| | N | 93 | 93 | 93 | 93 | 93 | 93 | 93 |
| XB23 | Pearson Correlation | ,082 | -,033 | 1 | ,316** | ,499** | ,384** | ,613** |
| | Sig. (2-tailed) | ,437 | ,754 | | ,002 | ,000 | ,000 | ,000 |
| | N | 93 | 93 | 93 | 93 | 93 | 93 | 93 |
| XB24 | Pearson Correlation | -,004 | ,148 | ,316** | 1 | ,656** | ,621** | ,747** |
| | Sig. (2-tailed) | ,973 | ,157 | ,002 | | ,000 | ,000 | ,000 |
| | N | 93 | 93 | 93 | 93 | 93 | 93 | 93 |
| XB25 | Pearson Correlation | ,022 | -,012 | ,499** | ,656** | 1 | ,749** | ,804** |
| | Sig. (2-tailed) | ,838 | ,907 | ,000 | ,000 | | ,000 | ,000 |
| | N | 93 | 93 | 93 | 93 | 93 | 93 | 93 |
| XB26 | Pearson Correlation | ,028 | ,081 | ,384** | ,621** | ,749** | 1 | ,789** |
| | Sig. (2-tailed) | ,787 | ,438 | ,000 | ,000 | ,000 | | ,000 |
| | N | 93 | 93 | 93 | 93 | 93 | 93 | 93 |
| Amount | Pearson Correlation | ,345** | ,348** | ,613** | ,747** | ,804** | ,789** | 1 |
| | Sig. (2-tailed) | ,001 | ,001 | ,000 | ,000 | ,000 | ,000 | |
| | N | 93 | 93 | 93 | 93 | 93 | 93 | 93 |

**. Correlation is significant at the 0.01 level (2-tailed).

**Table 9. Comparison of Test Results r Count with r Table of Importance Perceived Ease of Use XB2**

| Variable | r Count | r Table | Decision |
|---|---|---|---|
| XA21 | 0,345 | 0,206 | *Valid* |
| XA22 | 0,348 | 0,206 | *Valid* |
| XA23 | 0,613 | 0,206 | *Valid* |
| XA24 | 0,747 | 0,206 | *Valid* |
| XA25 | 0,804 | 0,206 | *Valid* |
| XA26 | 0,789 | 0,206 | *Valid* |

## 3.2. Reliability Test

Reliability is a tool for measuring a questionnaire which is an indicator of a variable or construct. A questionnaire is said to be reliable or reliable if a person's answer to a statement is consistent or stable over time. SPSS provides facilities for measuring reliability with the Cronbach Alpha statistical test.

### 3.2.1 Results Reliability Test of Importance Perceived Ease of Use XA1

**Table 10. Results Reliability Test of Importance Perceived Ease of Use XA1**

**Reliability Statistics**

| Cronbach's Alpha | N of Items |
|---|---|
| ,804 | 6 |

Cronbach's Alpha is an instrument reliability coefficient where the Perceived Ease of Use variable XA1 gets the Cronbach's alpha value is 0.804.

**3.2.2 Results Reliability Test of Importance Perceived Ease of Use XA2**

**Table 11. Results Reliability Test of Importance Perceived Ease of Use XA2**

**Reliability Statistics**

| Cronbach's Alpha | N of Items |
|---|---|
| ,784 | 6 |

Cronbach's Alpha is an instrument reliability coefficient where the Perceived Ease of Use variable XA2 gets the Cronbach's alpha value is 0,784.

**3.2.3 Results Reliability Test of Importance Perceived Ease of Use XB1**

**Table 12. Results Reliability Test of Importance Perceived Ease of Use XB1**

**Reliability Statistics**

| Cronbach's Alpha | N of Items |
|---|---|
| ,846 | 6 |

Cronbach's Alpha is an instrument reliability coefficient where the Perceived Ease of Use variable XB1 gets the Cronbach's alpha value is 0,846.

**3.2.4 Results Reliability Test of Importance Perceived Ease of Use XB2**

**Table 13. Results Reliability Test of Importance Perceived Ease of Use XB2**

**Reliability Statistics**

| Cronbach's Alpha | N of Items |
|---|---|
| ,659 | 6 |

Cronbach's Alpha is an instrument reliability coefficient where the Perceived Ease of Use variable XB2 gets the Cronbach's alpha value is 0,659.

From the test results above, it can be compared in the table below:

**Table 14. Comparison of the Reliability Test Results of Each Variable**

| Variable | *Croanbach's Alpha Value* | r Table | Information |
|---|---|---|---|
| Results Reliability Test of Importance Perceived Ease of Use XA1 | 0,804 | 0,60 | *Reliable* |
| Results Reliability Test of Importance Perceived Ease of Use XA2 | 0,784 | 0,60 | *Reliable* |
| Results Reliability Test of Importance Perceived Ease of Use XB1 | 0,846 | 0,60 | *Reliable* |

| | Results Reliability Test of Importance Perceived Ease of Use XB2 | 0,659 | 0,60 | *Reliable* |
|---|---|---|---|---|

The basis for making the reliability test decision is if the Croanbach's Alpha value is> 0.60 then the questionnaire or questionnaire is declared reliable or consistent. If the Croanbach's Alpha value <0.60 then the questionnaire or questionnaire is declared unreliable or inconsistent.

From the table above, it is known that the alpha value for the Perceived Ease of Use XA1 reliability test results is 0.804, the alpha value for the perceived importance level reliability test results (Perceived Ease of Use) XA2 is 0.784, the alpha value for the Perceived Ease of Use XB1 reliability test results of 0.846 and the alpha value for the Perceived Ease of Use XB1 performance reliability test results of 0.659. All variables show a value greater than r table which is equal to 0.60 so that all variables are Reliable.

### 3.3 Importance Performance Analysis (IPA)

After testing the validity and reliability of the research instrument and it was found that the instrument was valid and reliable, then the gap was analyzed between the Performance Level and the Importance Level of users to the quality of existing GoPay E-Wallet which is shown as follows:

**Table 15. Gap Analysis Importance and Performance E-Wallet GoPay User**

| No | Variable | Performance | Importance | Satisfaction Score |
|---|---|---|---|---|
| | **Perceived Ease of Use** | | | |
| 1. | The E-wallet Go-Pay is easy to learn | 3,8776 | 4,4694 | -0,59184 |
| 2. | Easy to control E-wallet Go-Pay. | 4,0102 | 4,4082 | -0,39796 |
| 3. | E-wallet Go-Pay is clear and understandable | 4,1020 | 4,4694 | -0,36735 |
| 4. | E-wallet Go-Pay flexible | 3,9082 | 4,4898 | -0,58163 |
| 5. | It is easy to use the E-Wallet Go-Pay to become skilled | 3,9694 | 4,4592 | -0,48980 |
| 6. | E-wallet Go-Pay easy to use | 3,9388 | 4,4490 | -0,51020 |
| | **Perceived Usefulness** | | | |
| 7. | I advise others to use the E-Wallet Go-Pay | 3,9490 | 4,0102 | -0,06122 |
| 8 | I advise my friends to use E-Wallet Go-Pay | 3,9898 | 4,1020 | -0,11224 |
| 9. | E-Wallet Go-Pay helps to improve my work | 4,0816 | 4,0816 | 0,00000 |
| 10. | E-Wallet Go-Pay helps increase my productivity | 4,0612 | 3,9592 | 0,10204 |
| 11. | E-Wallet Go-Pay helps to enhance my effectiveness | 4,0306 | 4,0306 | 0,00000 |
| 12. | In general the E-Wallet Go-Pay is useful for me | 4,0714 | 4,0408 | 0,03061 |

In the table it can be seen that the Performance Level column is generally lower than the Importance column, thus the satisfaction score column uses the formula Performance Level - Importance = Satisfaction Score. The application of the formula is applied to the satisfaction score column and it can be seen that the value in the column is predominantly negative, with the understanding that there are still variables tested that do not meet the respondent's expectations. In other words, it can be said that in general users are not yet satisfied with the quality of existing GoPay E-Wallet. There are many things related to the quality of GoPay E-Wallet that must be improved.

To find out the priority scale of improvements to E-Wallet GoPay, further analysis is carried out with the IPA (Importance Performance Analysis) tool where the items are mapped into a IPA graph which is divided into four quadrants as follows:

3. 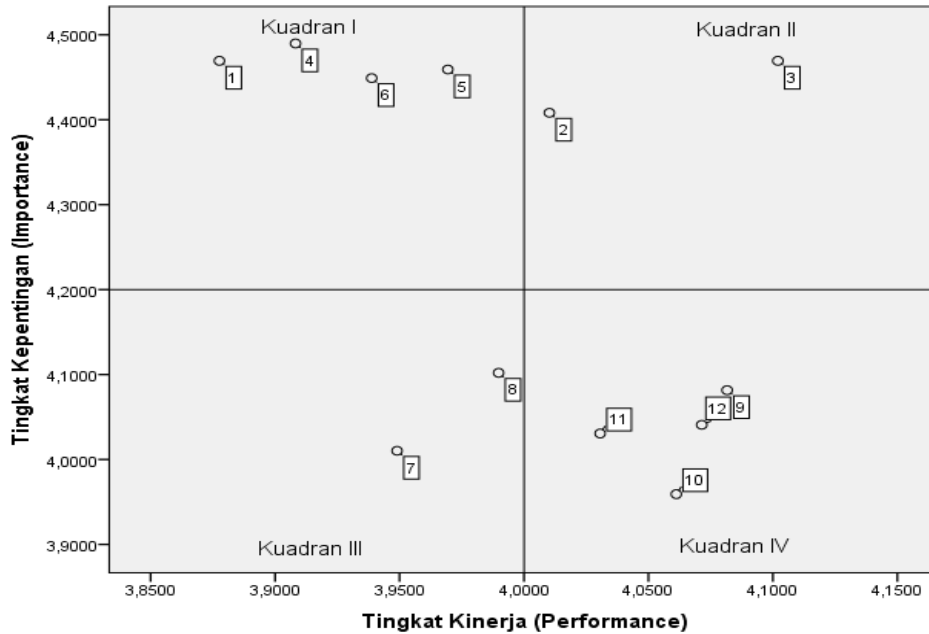

Figure IPA Graph

with the following priority scale

**Quadrant I**
Items included in this quadrant are the top priority for improving the quality of E-Wallet Go-Pay, which consists of:
[1] The E-Wallet Go-Pay is easy to learn.
[4] Go-Pay's e-wallet is flexible.
[5] Using the E-Wallet Go-Pay is easy to get skilled.
[6] The E-Wallet Go-Pay is easy to use.
The six items above are areas that are important according to user perceptions but have not met the expectations / expectations of the user so that they need to be corrected immediately to match the user's expectations.

**Quadrant II**
Items included in this quadrant are GoPay E-Wallet achievements or advantages that must be maintained because they meet user expectations, namely:
[2] Easy to control E-Wallet Go-Pay.
[3] The E-Wallet Go-Pay is clear and understandable.
The two items above are areas that are important according to user perceptions and are deemed to have met user expectations.

**Quadrant III**
Items included in this quadrant are a low priority scale for GoPay E-Wallet users because they are not considered important by users, namely:
[9] E-Wallet Go-Pay helps to improve my work.
[10] E-Wallet Go-Pay helps to increase my productivity.
[11] E-Wallet Go-Pay helps to enhance my effectiveness.
[12] In general the E-Wallet Go-Pay is useful for me.
The four items above are areas that are considered unimportant by the user so that they have low priority and can be ignored by the manager.

**Quadrant IV**
Items included in this quadrant are areas that are considered excessive because they are not considered important by the user but have high perceptions / performance, namely:
[7] I advise others to use the E-Wallet Go-Pay.
[8] I suggested to my friends to use the E-Wallet Go-Pay.
The two items above need to divert resources to a higher priority scale, namely quadrant I or quadrant II.

Based on the gap analysis, it is found that in general GoPay E-Wallet users are not satisfied with the current service quality. Based on the IPA analysis, the priority scale of GoPay E-Wallet quality improvement can be mapped where quadrant I is the highest priority scale according to the user's perspective: [1], [4], [5], and [6]. These three items must be upgraded immediately by the manager to meet user expectations. Areas of achievement or excellence for GoPay E-Wallet that must be maintained are in quadrant II, namely: [2] and [3]. From the explanation above, it can be concluded that in general the GoPay E-Wallet Service should be improved,

## 4. Conclusion

The test results on the validity test of all research instruments state that r Count is greater than the value of r table, it can be concluded that this research can be continued. The alpha value for the perceived importance level reliability test (Importance) XA1 perceived ease of use is 0.804, the alpha value for the perceived importance level reliability test (Importance) perceived ease of use (Perceived Ease of Use) XA2 is 0.784, The alpha value for the performance level reliability test results (Perceived Ease of Use) XB1 is 0.846 and the alpha value for the performance level reliability test results (Perceived Ease of Use) XB2 is 0.659. All variables show a value greater than r table which is equal to 0.60 so that all variables are reliable. All variables show a value greater than r table which is equal to 0.60 so that all variables are reliable.

Based on the gap analysis, it is found that in general users of GoPay E-Wallet quality are not satisfied with the quality of existing applications. Based on the IPA analysis, the priority scale of GoPay E-Wallet quality improvement can be mapped, where quadrant I is the highest priority scale according to the user's perspective, which are: [1], [4], [5], and [6]. The four items above are areas that are important according to user perceptions but have not met the expectations/ expectations of the user so they need to be corrected immediately so that they are in line with user expectations.Areas that are the achievements or advantages of GoPay E-Wallet that must be maintained because they meet user expectations, namely:
[2] Easy to control E-Wallet Go-Pay.
[3] The E-Wallet Go-Pay is clear and understandable.
The two items above are areas that are important according to user perceptions and are deemed to have met user expectations.

From the above explanation, it can be concluded that in general the GoPay E-Wallet Service must be improved, this shows that Users of the GoPay E-Wallet Service are not satisfied with the current Usability and Convenience.